\documentclass[preprint,aps, preprintnumbers, nofootinbib]{revtex4}
\usepackage[dvips]{graphics}

\newcommand{\lsim}[1]{
\setlength{\unitlength}{12pt}
\begin{picture}(1.4,1.)
\put(.7,-0.3){\makebox(0.0,1.)[t]{$<$}}
\put(.7,-0.3){\makebox(0.0,1.)[b]{$\sim$}}
\end{picture}#1}

\newcommand{\gsim}[2]{
\setlength{\unitlength}{12pt}
\begin{picture}(1.4,1.)
\put(.7,-0.3){\makebox(0.0,1.)[t]{$>$}}
\put(.7,-0.3){\makebox(0.0,1.)[b]{$\sim$}}
\end{picture}#2}

\begin{document}

\title{Effective field theory, large number of particle species, and
holography}

\author{R. Horvat}
\email{horvat@lei3.irb.hr}
\address{Rudjer Bo\v{s}kovi\'{c} Institute, P.O.B. 180, 10002 Zagreb, Croatia}

\begin{abstract}
An effective quantum field theory (QFT) with a manifest UV/IR connection, so
as to be valid for arbitrarily large volumes, can successfully be applied to
the cosmological dark energy problem as well as the cosmological constant (CC) 
problem. Motivated by recent approaches to the hierarchy problem, we
develop such a framework with a large number of particle species. When
applying to systems on the brink  of experiencing a sudden collapse to a 
black hole, we find that the entropy, unlike the total energy, now becomes
an increasing function of the number of field species. An internal
consistency of the theory is then used to infer the upper bound on the
number of particle species, showing consistency with the holographic
Bekenstein-Hawking bound. This may thus serve to fill in a large gap 
in entropy of any non-black hole configuration of matter and the black 
holes. In addition, when the bound is saturated the entanglement entropy
matches the black hole entropy, thus solving the multiplicity of species
problem. In a cosmological setting, the maximum allowable
number of species becomes a function of cosmological time, reaching its 
minimal value in a low-entropy post-reheating epoch.

\end{abstract}

\newpage

\maketitle

For an effective quantum field theory (QFT) in a box of size $L$ (providing an IR
cutoff) and with the UV cutoff $\Lambda $, the entropy scales extensively,
$S_{QFT} \sim L^3 \Lambda^3 $, and therefore there is always a sufficiently large 
volume
for which $S_{QFT}$ would exceed the absolute Bekenstein-Hawking bound $S_{BH}
\sim L^2 M_{Pl}^2$. Thus, considerations for the maximum possible entropy
suggest that ordinary QFT may not be valid for arbitrarily large volumes,
unless the UV and the IR cutoffs obey a constraint, $L \Lambda^3 \lsim
M_{Pl}^2$. \cite{1}. However, at saturation, this bound means that an
effective QFT should also be capable to describe systems containing 
black holes, since it necessarily includes many states with Schwarzschild 
radius much larger than the box size. There are however arguments for why an 
effective QFT
appears unlikely to
provide an adequate description of any system containing black holes
\cite{2, 3}. So, ordinary QFT may not be
valid for much smaller volumes, but would apply provided a more stringent
constraint, $L^{3/2} \Lambda^3 \lsim M_{Pl}^{3/2}$, is obeyed \cite{1}.

The above field-theoretical setup with the encoded holographic information
has recently triggered a novel variable CC approach, generically dubbed that
of `holographic dark energy' (HDE) \cite{4, 5, 6}. For the saturated case,
$\Lambda \sim L^{-1/2}$, $\Lambda $ gets depleted in an expanding universe
so that at present times the effective cosmological constant (CC) generated by
vacuum fluctuation (always dominated by UV modes) becomes so low that the
need for fine-tuning in the `old'  CC problem gets eliminated. Moreover,
with $L$ of order of the present Hubble radius, the CC energy density $\sim
L^{-2} M_{Pl}^2 $ becomes of the same order as the observed dark energy
of the universe \cite{7}. 

An approach to the hierarchy problem put forward a decade ago \cite{8}
demonstrates that the true UV cutoff can be made many orders of magnitude
smaller than the Planck mass $M_{Pl}$, provided proliferation of a large
number of quantum fields does occur in the theory. Besides the
higher-dimensional scenario of \cite{8}, hosting additional particles of the
Kaluza-Klein type, a similar scenario has appeared recently \cite{9, 10, 11}
in four dimensions, where the stability of the weak scale was explained
by postulating the existence of $N \sim 10^{32}$ 
gravitationally interacting species beyond the standard model. In a different 
context, looking at the renormalization-group running effects of $M_{Pl}$, 
the  same conclusion was reached in \cite{12}. Another, more natural
explanation for the weakness of gravity in particle physics requires a 
switch of statistic from Bose/Fermi to infinite one at high energies and no 
introduction of artificially large numbers \cite{13}. On nonperturbative grounds, a
cutoff $M_{Pl}/\sqrt{N}$ makes quantum entanglement universal \cite{11}, 
offering thus a resolution of the species problem in the physics of black 
holes.       

In the present paper, we examine a large-$N$ formulation of the effective 
QFT with UV/IR mixing underlying the saturated HDE models, i.e. 
obeying $L^{3/2}
\Lambda^3 \simeq M_{Pl}^{3/2}$. Note that the present size of the universe
is large enough to reduce the  UV cutoff down even to the dark-energy scale of
$10^{-3}$ eV. Hence there is  no need  to introduce a large number $N$ of
particle species  to reduce $\Lambda $ any further, the motivation here
being different than in  
models motivated by a stabilization of the weak scale. Next 
we list a few obvious benefits of the large-$N$ formulation of an effective
QFT with the proposed relationship between the UV and the IR cutoffs:(i) For
systems on the verge of gravitational collapse we find their entropy to
scale as $N^{1/4}L^{3/2}M_{Pl}^{3/2}$, realizing thus a possibility to
complete a large gap in entropy between those systems, $L^{3/2}M_{Pl}^{3/2}$
\cite{2} and that of black
holes, $L^2 M_{Pl}^2$; (ii) An internal consistency of the theory yields a
bound, $N_{max} \simeq L^2 M_{Pl}^2 $, and for the particular model 
we can trace $L$ to the earliest moments in the history of the universe to
obtain a minimum value for $N_{max}$; (iii) For systems with $N = N_{max}$,
when the access to the inside region of the system becomes impossible for the
outside observer, we find that the entanglement entropy (scaling up with $N$)
matches the black hole entropy (a universal quantity independent of the
black-hole past history 
that should not depend
on $N$). We note that a large-$N$ scenario with the UV/IR connection in 
higher-dimensional settings was also recently discussed \cite{14}.       

We begin our considerations by recapitulating the scenario of
Cohen, Kaplan and Nelson  
in a different manner, such as to allow us to lay down an extra feature 
not exposed in \cite{1}, which turns out to be 
crucial for our arguments: a lower bound on the 
QFT energy density $\rho_{\Lambda}$ (or equivalently on $\Lambda $). 
\footnote{Obviously, $\rho_{\Lambda}$ is the energy density corresponding to
a zero-point energy and the cutoff $\Lambda $, being proportional to
$\Lambda^4 $ ($\Lambda
\gsim \; m$) or $m\Lambda^3$ ($\Lambda \lsim m$), where $m$ is the mass of the
QFT field. Throughout the paper we  shall explore the consequences of the
former choice which turn out to be more interesting, and mention the latter
case only at the end of the paper.} For that purpose also the
Bekenstein bound $S_B $ \cite{15} needs to be  invoked. For a
macroscopic system in which self-gravitation effects can be disregarded, the 
Bekenstein bound is given by a product of the energy and the linear size of
the system, $EL$. In the context of the effective QFT it therefore becomes
proportional to $\rho_{\Lambda}^4 L^4 $. It is convenient to use this entropic
bound to derive a lower bound on the energy density $\rho_{\Lambda}$ 
since $S_B $ is more extensive than $S_{QFT}$. The obvious hierarchy
between $S_{QFT} \sim L^3 \Lambda^3 $ and the entropic bounds, 
$S_B \sim \rho_{\Lambda} L^4 $ and
$S_{BH} \sim L^2 M_{Pl}^2$,   
\begin{equation}
S_{QFT} \leq S_B \leq S_{BH} \;,
\end{equation}
yields $\rho_{\Lambda}$ which is constrained from both above and
below:\footnote{We note that the entropic bound of type (1) prevents
to assign the CC a zero value.}
\begin{equation}
\Lambda^3 L^{-1} \leq \rho_{\Lambda} \leq L^{-2} M_{Pl} \;.
\end{equation}
From (1) and (2) one sees that the concept of HDE emerges whenever 
$S_B \leq S_{BH}$; i.e.,
for a weakly gravitating system. This requirement 
automatically prevents formations of black holes, as the Bekenstein bound, in spite
of its original connection with black hole physics, does not involve the
Newton constant. The most commonly used saturated models do obey $S_B \simeq
S_{BH}$; this requirement  
brings a system on the verge of gravitational collapse since then
$L \simeq L_S $, with $L_S \sim M_{PL}^{-2} \rho_{\Lambda} L^3$ being the
Schwarzschild radius. The only remaining bound is then the lower bound on
$\rho_{\Lambda}$, which for $\rho_{\Lambda} \sim \Lambda^4 $ simply becomes
$\Lambda \gsim \; L^{-1}$. Below we are going to consider this bound as a
consistency condition which will allow us to set the upper bound on $N$.

Now we are going to implement the new $N$ degrees of freedom  into the setup
described by (1) and (2). The main observation is that while both $S_{QFT}$
and $S_B $ scale up with the number $N$ of the species, $S_{BH}$ stays a 
universal quantity that does not depend on $N$. The setup underlying the
saturated HDE models with $N >> 1$ thus becomes
\begin{equation}
N S_{QFT} \leq N S_B \simeq S_{BH} \;.
\end{equation}
From (3) one readily sees that the lower bound $\Lambda \gsim \; L^{-1}$
remains unchanged. What we  would like to find out is the upper bound on
the entropy in QFT, $N \Lambda^3 L^3 $. From $N S_B \simeq S_{BH}$, i.e., 
\begin{equation}
N \Lambda^4 L^4 \simeq L^2 M_{PL}^2 \;,
\end{equation}
one finds that the upper bound on $N S_{QFT}$ becomes $N$-dependent \footnote{
In contrast, the upper bound for the total energy $N \Lambda^4 L^3 $ stays
$N$-independent. Also, (4) and (5) still continue to describe systems on the
verge of gravitational collapse since now the Schwarzschild radius scales 
up with $N$.},
\begin{equation}
N \Lambda^3 L^3 \simeq N^{1/4}  L^{3/2} M_{PL}^{3/2} \;.
\end{equation}
Extracting $\Lambda /L^{-1}$ from (5) and setting $\Lambda /L^{-1} \gsim \;
1$, 
\begin{equation}
\frac{\Lambda }{L^{-1}} \simeq N^{-1/4} L^{1/2} M_{PL}^{1/2} \gsim \; 1 \;,
\end{equation}
one obtains the upper bound for $N$ in the form
\begin{equation}
N \lsim L^2 M_{PL}^2 \;,
\end{equation}
being of order of the Bekenstein-Hawking entropy itself. From (5) we see that 
the gap between systems on the verge of gravitational
collapse having $S \sim L^{3/2}M_{Pl}^{3/2}$ \cite{2} and black holes begins 
to populate when $N$ is increasing \footnote{Another resolution of this
problem involves curved space configurations called {\it monsters} \cite{16}}. 
When the
bound on $N$ (7) begins to saturate, a (normal) system begins to 
sustain a black
hole entropy. Note that (7) corresponds to a loose bound in the
late-time universe, $N \lsim 10^{122}$. Also, at saturation, $L \simeq
\Lambda^{-1}$, Eq. (7) does reproduce the gravity cutoff $M_{Pl}/\sqrt{N}$
obtained in the scenario \cite{9, 10, 11}.  

We note the ratio $\Lambda /L^{-1}$ in (6) as the increasing function 
of the IR
cutoff $L$, which means that in an expanding universe the upper bound on $N$
can be strengthened considerably, provided we have some knowledge on the behavior
of $L$ in the past. For that purpose we have to resort to a particular
model. This makes the upper bound on $N$ model-dependent. For the sake of
illustration, we consider the popular Li's model \cite{5}. This model
belongs to a class of 
noninteracting and saturated HDE models, with a choice for $L$ in the form of
the future event horizon,
\begin{equation}
d_{E} = a \int_{a}^{\infty } \frac{da}{a^2 H} \;,
\end{equation}
with $a$ being a scale factor. Furthermore, we assume that our vacuum energy
$\rho_{\Lambda}$ is not responsible for the early-time inflation, and that
all particle species came into being when early vacuum energy density decays
into matter, in the process of reheating (see e.g., \cite{17}). Ignoring
subtle details of reheating, we assume an instantaneous process, occurring at
$T_{reh}$. This amounts to knowing the behavior of $\rho_{\Lambda} \simeq
L^{-2} M_{PL}^2 $ during the radiation-dominated era, in which $\rho_{\Lambda}$
occupies only a tiny fraction of the total energy density. In a
two-component universe $\rho_{\Lambda}$ evolution is governed 
by \cite{5, 18}
\begin{equation}
\Omega_{\Lambda }^{'} = \Omega_{\Lambda }^2 ( 1 - \Omega_{\Lambda }) \left
[\frac{1}{\Omega_{\Lambda }} + \frac{2}{c \sqrt{\Omega_{\Lambda }}} \right ]
\;,
\end{equation}
where the prime denotes the derivative with respect to $lna $. In (9)
$\Omega_{\Lambda } = \rho_{\Lambda }/\rho_{crit} $, where $\rho_{crit}$ is the
critical density and $\rho_{\Lambda } $ was parametrized as $\rho_{\Lambda }
=(3/8 \pi ) c^2 M_{Pl}^2 L^{-2}$. With $ \Omega_{\Lambda } << 1$, $ c \simeq
1$ and  $\rho_{crit} \simeq  \rho_{rad }$ we obtain
\begin{equation}
\rho_{\Lambda } \simeq \rho_{rad 0} a^{-3} \;,
\end{equation}
where $\rho_{rad 0}$  denotes the radiation energy density at the present
time. This in turn determines $L(a)$ as
\begin{equation}
L(a) \simeq M_{Pl} \; \rho_{rad 0}^{-1/2} \; a^{3/2} \;.
\end{equation}
Equipped with these relationships and $T \sim a^{-1}$, we obtain a final
expression for $N_{max} = L^2 M_{Pl}^2 $ as
\begin{equation}
N_{max} \simeq M_{Pl}^4 \; \rho_{rad 0}^{-1} \; (T_{reh}/T_0 )^{-3} \;,
\end{equation}
where $T_0 $ is the present temperature of the universe. In supergravity
theories if the reheating temperature after inflation is too high one 
inevitably overproduce gravitinos. Plugging in the relevant numbers with
$T_{reh}^{max} \simeq 10^7 GeV$ \cite{19} as to avoid troubles with
the overproduction of gravitinos, one gets $N_{max} \simeq 10^{68}$, a
considerably stringent requirement than $10^{122}$. The bound $N_{max}
\simeq 10^{32}$ \cite{9, 10, 11, 12}, obtained in QFTs without the holographic
constraint, by noting that we have not seen any strong gravity in the
particle collisions, is still more stringent. We stress once again
that the bound obtained here is quite model-dependent. In addition, a 
higher $T_{reh}$
could reduce $N_{max}$ considerably. Also, if $\rho_{\Lambda }
$ is to play any role in early-time inflation, one expects much severe
constraints on $N_{max}$. Still, our bound is much less than the entropy of
the CMB photons or relic neutrinos in the present universe $(\sim 10^{88})$.  

Now we show that when the bound (7) is saturated, the entanglement entropy
$S_{ent}$ \cite{20} computed in the proposed QFT setup can be the origin of 
black hole
entropy. When the bound (7) is not saturated, an observer outside of the box 
of size $L$ has (at least theoretically) an unlimited access to the 
interior of the box.
Consequently, the entanglement entropy, measuring quantum-mechanical
correlations between the box and the space outside of the box, is 
zero. However,
at saturation of (7) the physical horizon forms, and consequently 
an outside observer lacks any information about the interior of the box.
Thus, both the entanglement and the black hole entropy then become nonzero
\footnote{A small nonvanishing entanglement entropy emerges even for systems
with $L \simeq L_S $ and $N \lsim N_{max}$, as for the systems with 
artificially created horizons \cite{21}. However, the only possible way to
physically prevent the access to a part of the system is to put a closed
surface separating the two subsystems on the event horizon.}.
A pressing problem in identification of $S_{BH}$ with $S_{ent}$  is the
multiplicity of species problem \cite{22}. Incidently, $S_{ent}$ should
depend on $N$, while $S_{BH}$ lacks any information about number of species.
We show below that the proposed UV/IR mixing, together with the bound (7), 
easily resolves this dilemma.
In dealing with an overall pure state, $S_{ent}$ should behave nonextensively, i.e.,
should depend only on the surface $A(L) \sim L^2$ separating the box from the
rest. On the other hand, quantum correlations between the subsystems in
local QFT are taken care by the UV cutoff. Hence we have
\begin{equation}
S_{ent}(L) \simeq \Lambda^2 \; A(L) \;.
\end{equation}
By invoking (4) we have
\begin{equation}
\Lambda^2 \simeq N^{-1/2}\; L^{-1} \; M_{Pl} \;.
\end{equation}
Noting that $S_{ent}$ should scale up with $N$ one has
\begin{equation}
S_{ent}(L, N) \simeq N \times (N^{-1/2}\; L^{-1} \; M_{Pl} \;) \times A(L)
\simeq N^{1/2}  \; M_{Pl}  \; L  \;.
\end{equation}
With $N_{max} = L^2 M_{Pl}^2$, one immediately arrives at 
\begin{equation}
S_{ent}(L, N_{max}) \simeq M_{Pl}^2 L^2 \simeq S_{BH} \;.
\end{equation}
Thus, we have seen how the proposed UV/IR mixing,
together with the bound on the number of species, settles the problem of
species. 

In considering the  second option, $\rho_{\Lambda } \sim m\Lambda^3$ 
($\Lambda \lsim m$), one should replace the consistency relation $\Lambda
\gsim \; L^{-1}$ with $m \gsim \; L^{-1}$, which is nothing but a trivial
statement about encompassment of the modes within the box. So, $N$ cannot be
restricted before $\Lambda \gsim \; L^{-1}$ is imposed by hand. This is, of
course, a quite plausible constraint for any QFT. The upper bound for
$N L^3 \Lambda^3 $ then becomes $m^{-1} L M_{Pl}^2 $, which means that the gap
between normal systems and black holes can never be fully populated. The
same is also true for the entanglement entropy. This makes this case less
interesting.  

In conclusion, we have developed a promising QFT setup 
of Cohen, Kaplan and Nelson with a large-$N$ new
degrees of freedom. The holographic ingredient implemented via the specific
UV/IR mixing  makes the setup valid in an arbitrarily large volume, such that
successful application both to particle physics and cosmology becomes possible.
By using thermodynamics with large $N$,  we have shown that it is possible
to  bridge a gap in entropy between the systems on the verge of gravitational 
collapse and the black holes themselves. Drawing on the internal consistency of
the theory and cosmological evolution of the IR cutoff, we have obtained the
upper bound for the number of particle species $N$. Finally, a resolution of
the species problem comes out naturally due to the proposed UV/IR relationship.

{\bf Acknowledgment. } This work was supported by the Ministry of Science,
Education and Sport
of the Republic of Croatia under contract No. 098-0982887-2872.

\end{document}